\documentstyle[12pt]{article}
\begin{document}
\title{SHEARING EXPANSIONFREE SPHERICAL ANISOTROPIC FLUID EVOLUTION}
\author{L. Herrera$^1$\thanks{e-mail: laherrera@cantv.net.ve}, N. O. Santos$^{2,3}$
\thanks{e-mail: N.O.Santos@qmul.ac.uk}
and Anzhong Wang$^{4,5}$\thanks{e-mail: anzhong\_wang@baylor.edu}\\
{\small $^1$Escuela de F\'{\i}sica, Facultad de Ciencias,}\\
{\small Universidad Central de Venezuela, Caracas, Venezuela.}\\
{\small $^2$School of Mathematical Sciences, Queen Mary,}\\
{\small University of London, London E1 4NS, UK.}\\
{\small $^3$Laborat\'orio Nacional de Computa\c{c}\~ao Cient\'{\i}fica,
25651-070 Petr\'opolis RJ, Brazil.}\\
{\small $^4$GCAP-CASPER, Department of Physics,}\\
{\small Baylor University, Waco, Texas 76798-7316, USA.}\\
{\small $^5$Department of Theoretical Physics, State University of Rio de Janeiro, RJ, Brazil.}}
\maketitle

\begin{abstract}

\end{abstract}
Spherically symmetric expansionfree distributions are systematically studied. The whole set of field equations and junction conditions are presented for a general  distribution of  dissipative anisotropic fluid  (principal stresses unequal), and the expansionfree condition is integrated. In order to understand the physical meaning of expansionfree motion, two different definitions  for the  radial velocity of a fluid element  are discussed. It is shown that the appearance of a cavity is inevitable in the expansionfree evolution. The  nondissipative  case is  considered in detail and the Skripkin model is 
recovered.
\newpage
\section{Introduction}
The problem of general relativistic gravitational collapse of massive stars has attracted
the attention of researchers for many years, starting with   the seminal paper by Oppenheimer
and Snyder \cite{Opp}. The motivation for such interest is easily
understood: the gravitational collapse of massive stars represents one of
the few observable phenomena, where general relativity is expected to
play a relevant role. Ever since that work, much was written by
researchers trying to provide models of evolving gravitating spheres (see \cite{May} and references therein).
However this endeavour proved to be difficult. Different
kinds of obstacles appearing, depending on the approach adopted for the
modelling.

Thus, numerical methods  allow for
considering more realistic equations of state, but the obtained results,
in general, are restricted and highly model dependent. Also, specific difficulties, associated to numerical solutions of partial
differential equations in presence of shocks, complicate further the
problem. Therefore it seems useful to consider nonstatic models which are relatively simple to analyze but still contain some of the essential features of a realistic situation. For doing so we need to appeal  to a simple equation of state and/or to  additional physically meaningful
heuristic assumptions. In this work, we shall assume the fluid to be expansionfree.

As is well known, the motion of  a fluid may be characterized by the four acceleration vector ($a^\alpha$), the shear tensor ($\sigma_{\alpha\beta}$), the expansion scalar ($\Theta$) and  the vorticity tensor (which vanishes in the spherically symmetric  case).
The relevance of the shear tensor in the evolution of selfgravitating systems and the consequences emerging from its vanishing has been brought out by many authors (see  \cite{SFCF} and references therein).

In this work we  shall study the properties of an expansionfree  spherically symmetric selfgravitating fluid.

Since the expansion scalar describes the rate of change  of   small volumes of the fluid, it  is intuitively clear that the evolution of an expansionfree  spherically symmetric distribution should necessarily imply the formation of a vacuum cavity  within the distribution (see a more rigorous argument on this in section V).
Thus, in the case of an overall expansion, the increase in  volume due to the increasing  area of  the external boundary surface must be compensated with  the increase of the   area of the internal boundary surface (delimiting the cavity) in order to keep $\Theta$ vanishing. The argument in the case of collapse is similar.

For sake of generality we shall start our discussion by  considering an anisotropic  dissipative viscous fluid (arguments to justify such kind of fluid distributions may found in \cite{matter}-\cite{matter2}  and references therein). For this kind of distribution we shall write the field equations, the junction conditions, at, both, the inner and the external boundary surface (section II), as well as  the dynamical equations (section III). Next we shall integrate the expansionfree condition and find the general form of the metric for the anisotropic dissipative viscous fluid (section IV).

In order to understand better the physical meaning of the expansionfree motion, we shall discuss two different definitions of  radial  velocity of a fluid element, in terms of which both the expansion and the shear can be expressed (section V).

We shall next consider the nondissipative case, and  we shall specialize further to the isotropic fluid. In this latter case we shall recover as a particular example the Skripkin model \cite{Skripkin},  assuming the energy density to be constant (section VI).

Finally a discussion on the results is presented in the last section.

\section{The  energy-momentum tensor, the field equations and the junction conditions}
We consider a spherically symmetric distribution  of collapsing
fluid, bounded by a spherical surface $\Sigma^{(e)}$. The fluid is
assumed to be locally anisotropic (principal stresses unequal) and undergoing dissipation in the
form of heat flow (to model dissipation in the diffusion approximation), null radiation (to model dissipation in the free streaming approximation) and shearing
viscosity.

Choosing comoving coordinates inside $\Sigma^{(e)}$, the general
interior metric can be written
\begin{equation}
ds^2_-=-A^2dt^2+B^2dr^2+R^2(d\theta^2+\sin^2\theta d\phi^2),
\label{1}
\end{equation}
where $A$, $B$ and $R$ are functions of $t$ and $r$ and are assumed
positive. We number the coordinates $x^0=t$, $x^1=r$, $x^2=\theta$
and $x^3=\phi$. Observe that $A$ and $B$ are dimensionless, whereas $R$ has the same dimension as $r$.

The matter energy-momentum $T_{\alpha\beta}^-$ inside $\Sigma^{(e)}$
has the form
\begin{equation}
T_{\alpha\beta}^-=(\mu +
P_{\perp})V_{\alpha}V_{\beta}+P_{\perp}g_{\alpha\beta}+(P_r-P_{\perp})\chi_{
\alpha}\chi_{\beta}+q_{\alpha}V_{\beta}+V_{\alpha}q_{\beta}+
\epsilon l_{\alpha}l_{\beta}-2\eta\sigma_{\alpha\beta}, \label{3}
\end{equation}
where $\mu$ is the energy density, $P_r$ the radial pressure,
$P_{\perp}$ the tangential pressure, $q^{\alpha}$ the heat flux,
$\epsilon$ the energy density of the null fluid describing dissipation in the free streaming approximation, $\eta$ the coefficient of
shear viscosity, $V^{\alpha}$ the four velocity of the fluid,
$\chi^{\alpha}$ a unit four vector along the radial direction
and $l^{\alpha}$ a radial null four vector. These quantities
satisfy
\begin{equation}
V^{\alpha}V_{\alpha}=-1, \;\; V^{\alpha}q_{\alpha}=0, \;\; \chi^{\alpha}\chi_{\alpha}=1, \;\;
\chi^{\alpha}V_{\alpha}=0, \;\; l^{\alpha}V_{\alpha}=-1, \;\; l^{\alpha}l_{\alpha}=0.
\end{equation}
The acceleration $a_{\alpha}$ and the expansion $\Theta$ of the fluid are
given by
\begin{equation}
a_{\alpha}=V_{\alpha ;\beta}V^{\beta}, \;\;
\Theta={V^{\alpha}}_{;\alpha}. \label{4b}
\end{equation}
and its  shear $\sigma_{\alpha\beta}$ by
\begin{equation}
\sigma_{\alpha\beta}=V_{(\alpha
;\beta)}+a_{(\alpha}V_{\beta)}-\frac{1}{3}\Theta(g_{\alpha\beta}+V_{\alpha}V
_{\beta}),
\label{4a}
\end{equation}

We do not explicitly add bulk viscosity to the system because it
can be absorbed into the radial and tangential pressures, $P_r$ and
$P_{\perp}$, of the
collapsing fluid \cite{Chan}.

Since we assumed the metric (\ref{1}) comoving then
\begin{equation}
V^{\alpha}=A^{-1}\delta_0^{\alpha}, \;\;
q^{\alpha}=qB^{-1}\delta^{\alpha}_1, \;\;
l^{\alpha}=A^{-1}\delta^{\alpha}_0+B^{-1}\delta^{\alpha}_1, \;\;
\chi^{\alpha}=B^{-1}\delta^{\alpha}_1, \label{5}
\end{equation}
where $q$ is a function of $t$ and $r$.

From  (\ref{4b}) with (\ref{5}) we have for the  acceleration and its scalar $a$,
\begin{equation}
a_1=\frac{A^{\prime}}{A}, \;\; a^2=a^{\alpha}a_{\alpha}=\left(\frac{A^{\prime}}{AB}\right)^2, \label{5c}
\end{equation}
and for the expansion
\begin{equation}
\Theta=\frac{1}{A}\left(\frac{\dot{B}}{B}+2\frac{\dot{R}}{R}\right),
\label{5c1}
\end{equation}
where the  prime stands for $r$
differentiation and the dot stands for differentiation with respect to $t$.
With (\ref{5}) we obtain
for the shear (\ref{4a}) its non zero components
\begin{equation}
\sigma_{11}=\frac{2}{3}B^2\sigma, \;\;
\sigma_{22}=\frac{\sigma_{33}}{\sin^2\theta}=-\frac{1}{3}R^2\sigma,
 \label{5a}
\end{equation}
and its scalar
\begin{equation}
\sigma^{\alpha\beta}\sigma_{\alpha\beta}=\frac{2}{3}\sigma^2,
\label{5b}
\end{equation}
where
\begin{equation}
\sigma=\frac{1}{A}\left(\frac{\dot{B}}{B}-\frac{\dot{R}}{R}\right).\label{5b1}
\end{equation}

\subsection{The Einstein equations}
Einstein's field equations for the interior spacetime (\ref{1}) are given by
\begin{equation}
G_{\alpha\beta}^-=8\pi T_{\alpha\beta}^-,
\label{2}
\end{equation}
and its non zero components
with (\ref{1}), (\ref{3}) and (\ref{5})
become
\begin{eqnarray}
8\pi T_{00}^-=8\pi(\mu+\epsilon)A^2
=\left(2\frac{\dot{B}}{B}+\frac{\dot{R}}{R}\right)\frac{\dot{R}}{R}\nonumber\\
-\left(\frac{A}{B}\right)^2\left[2\frac{R^{\prime\prime}}{R}+\left(\frac{R^{\prime}}{R}\right)^2
-2\frac{B^{\prime}}{B}\frac{R^{\prime}}{R}-\left(\frac{B}{R}\right)^2\right],
\label{12} \\
8\pi T_{01}^-=-8\pi(q+\epsilon)AB
=-2\left(\frac{{\dot R}^{\prime}}{R}
-\frac{\dot B}{B}\frac{R^{\prime}}{R}-\frac{\dot
R}{R}\frac{A^{\prime}}{A}\right),
\label{13} \\
8\pi T_{11}^-=8\pi
\left(P_r+\epsilon-\frac{4}{3}\eta\sigma\right)B^2 \nonumber\\
=-\left(\frac{B}{A}\right)^2\left[2\frac{\ddot{R}}{R}-\left(2\frac{\dot A}{A}-\frac{\dot{R}}{R}\right)
\frac{\dot R}{R}\right]\nonumber\\
+\left(2\frac{A^{\prime}}{A}+\frac{R^{\prime}}{R}\right)\frac{R^{\prime}}{R}-\left(\frac{B}{R}\right)^2,
\label{14} \\
8\pi T_{22}^-=\frac{8\pi}{\sin^2\theta}T_{33}^-=8\pi\left(P_{\perp}+\frac{2}{3}\eta\sigma\right)R^2\nonumber \\
=-\left(\frac{R}{A}\right)^2\left[\frac{\ddot{B}}{B}+\frac{\ddot{R}}{R}
-\frac{\dot{A}}{A}\left(\frac{\dot{B}}{B}+\frac{\dot{R}}{R}\right)
+\frac{\dot{B}}{B}\frac{\dot{R}}{R}\right]\nonumber\\
+\left(\frac{R}{B}\right)^2\left[\frac{A^{\prime\prime}}{A}
+\frac{R^{\prime\prime}}{R}-\frac{A^{\prime}}{A}\frac{B^{\prime}}{B}
+\left(\frac{A^{\prime}}{A}-\frac{B^{\prime}}{B}\right)\frac{R^{\prime}}{R}\right].\label{15}
\end{eqnarray}
The component (\ref{13}) can be rewritten with (\ref{5c1}) and
(\ref{5b}) as
\begin{equation}
4\pi(q+\epsilon)B=\frac{1}{3}(\Theta-\sigma)^{\prime}
-\sigma\frac{R^{\prime}}{R}.\label{17a}
\end{equation}

Next, the mass function $m(t,r)$ introduced by Misner and Sharp
\cite{Misner} (see also \cite{Cahill}) reads
\begin{equation}
m=\frac{R^3}{2}{R_{23}}^{23}
=\frac{R}{2}\left[\left(\frac{\dot R}{A}\right)^2-\left(\frac{R^{\prime}}{B}\right)^2+1\right],
 \label{17masa}
\end{equation}

Thus in the most general case (locally anisotropic and dissipative) we have available four field equations (\ref{12}--\ref{15}) for  eight variables, namely  $A$, $B$, $R$, $\mu$, $P_r$, $P_{_\perp}$, $\epsilon$ and  $q$. Since we are going to consider expansionfree systems we have the additional condition $\Theta=0$. Evidently, in order to find specific models (to close the system of equations) we need to provide additional information, which could be given in the form of constitutive equations for $q$ and $\epsilon$, and equations of state for both pressures.

\subsection{The exterior spacetime and junction conditions}
Outside $\Sigma^{(e)}$ we assume we have the Vaidya
spacetime (i.e.\ we assume all outgoing radiation is massless),
described by
\begin{equation}
ds^2=-\left[1-\frac{2M(v)}{r}\right]dv^2-2drdv+r^2(d\theta^2
+\sin^2\theta
d\phi^2) \label{1int},
\end{equation}
where $M(v)$  denotes the total mass,
and  $v$ is the retarded time.

The matching of the full nonadiabatic sphere  (including viscosity) to
the Vaidya spacetime, on the surface $r=r_{\Sigma^{(e)}}=$ constant, was discussed in
\cite{chan1}. From the continuity of the first and second differential forms it follows (see \cite{chan1} for details)
\begin{equation}
m(t,r)\stackrel{\Sigma^{(e)}}{=}M(v), \label{junction1}
\end{equation}
and
\begin{eqnarray}
2\left(\frac{{\dot R}^{\prime}}{R}-\frac{\dot B}{B}\frac{R^{\prime}}{R}-\frac{\dot R}{R}\frac{A^{\prime}}{A}\right)
\nonumber\\
\stackrel{\Sigma^{(e)}}{=}-\frac{B}{A}\left[2\frac{\ddot R}{R}
-\left(2\frac{\dot A}{A}
-\frac{\dot R}{R}\right)\frac{\dot R}{R}\right]+\frac{A}{B}\left[\left(2\frac{A^{\prime}}{A}
+\frac{R^{\prime}}{R}\right)\frac{R^{\prime}}{R}-\left(\frac{B}{R}\right)^2\right],
\label{j2}
\end{eqnarray}
where $\stackrel{\Sigma^{(e)}}{=}$ means that both sides of the equation
are evaluated on $\Sigma^{(e)}$ (observe a misprint in eq.(40) in \cite{chan1} and a slight difference in notation).

Comparing (\ref{j2}) with  (\ref{13}) and (\ref{14}) one obtains
\begin{equation}
q\stackrel{\Sigma^{(e)}}{=}P_r-\frac{4}{3}\eta \sigma,\label{j3}
\end{equation}
Thus   the matching of
(\ref{1})  and (\ref{1int}) on $\Sigma^{(e)}$ implies (\ref{junction1}) and  (\ref{j3}),
which reduces to equation (41) in \cite{chan1} with the appropriate change in notation. Observe a misprint in equation (27) in \cite{matter1} (the $\sigma$ appearing there is the one defined in \cite{chan1}, which is $-1/3$ of the one used here and in \cite{matter1}).

As we mentioned in the introduction, the expansionfree models  present an internal vacuum cavity. If we call $\Sigma^{(i)} $ the boundary surface between the cavity and the fluid, then the matching of the Minkowski spacetime within the cavity to the fluid distribution, implies
\begin{equation}
m(t,r)\stackrel{\Sigma^{(i)}}{=}0, \label{junction1i}
\end{equation}
\begin{equation}
q\stackrel{\Sigma^{(i)}}{=}P_r-\frac{4}{3}\eta \sigma.\label{j3i}
\end{equation}
\section{Dynamical equations}
To study the dynamical properties of the system, let us  introduce,
following Misner and Sharp \cite{Misner}, the proper time derivative $D_T$
given by
\begin{equation}
D_T=\frac{1}{A}\frac{\partial}{\partial t}, \label{16}
\end{equation}
and the proper radial derivative $D_R$,
\begin{equation}
D_R=\frac{1}{R^{\prime}}\frac{\partial}{\partial r}, \label{23a}
\end{equation}
where $R$ defines the areal radius of a spherical surface inside $\Sigma^{(e)}$ ( as
measured from its area).

Using (\ref{16}) we can define the velocity $U$ of the collapsing
fluid (for another definition of velocity see section 6) as the variation of the areal radius with respect to proper time, i.e.\
\begin{equation}
U=D_TR<0 \;\; \mbox{(in the case of collapse)}. \label{19}
\end{equation}
Then (\ref{17masa}) can be rewritten as
\begin{equation}
E \equiv \frac{R^{\prime}}{B}=\left(1+U^2-\frac{2m}{R}\right)^{1/2}.
\label{20x}
\end{equation}
With (\ref{23a}) we can express (\ref{17a}) as
\begin{equation}
4\pi(q+\epsilon)=E\left[\frac{1}{3}D_R(\Theta-\sigma)
-\frac{\sigma}{R}\right].\label{21a}
\end{equation}

Using (\ref{12})-(\ref{14}) with (\ref{16}) and (\ref{23a}) we obtain from
(\ref{17masa})
\begin{eqnarray}
D_Tm=-4\pi\left[\left
(P_r+\epsilon-\frac{4}{3}\eta\sigma\right)U+(q+\epsilon)E\right]R^2,
\label{22Dt}
\end{eqnarray}
and
\begin{eqnarray}
D_Rm=4\pi\left[\mu+\epsilon+(q+\epsilon)\frac{U}{E}\right]R^2,
\label{27Dr}
\end{eqnarray}
which implies
\begin{equation}
m=4\pi\int^{R}_{0}\left[\mu +
\epsilon+(q+\epsilon)\frac{U}{E}\right]R^2dR \label{27intcopy}
\end{equation}
(assuming a regular centre to the distribution, so $m(0)=0$).

Expression  (\ref{22Dt}) describes the rate of variation of the
total energy inside a surface of areal radius $R$. On the right hand
side of (\ref{22Dt}), $(P_r+\epsilon-4\eta\sigma/3)U$ (in the case
of collapse $U<0$) increases the energy inside $R$ through the
rate of work being done by the ``effective'' radial pressure
$P_r-4\eta\sigma/3$ and the radiation pressure $\epsilon$. Clearly
here the heat flux $q$ does not appear since there is no
pressure associated with the diffusion process. The second term
$-(q+\epsilon)E$ is the matter energy leaving the spherical
surface.

Equation (\ref{27Dr}) shows how the total energy enclosed varies between
neighboring spherical surfaces inside the
fluid distribution.
The first term on the right hand side of (\ref{27Dr}), $\mu+\epsilon$, is due
to the energy density of the fluid element plus the energy density of the
null fluid describing dissipation
in the free streaming approximation. The second term,
$(q+\epsilon)U/E$ is negative (in the case of collapse) and measures the
outflow of heat and radiation.

The non trivial components of the Bianchi identities, $T^{-\alpha\beta}_{;\beta}=0$, from (\ref{2}) yield
\begin{eqnarray}
T^{-\alpha\beta}_{;\beta}V_{\alpha}=-\frac{1}{A}\left[(\mu+\epsilon)\dot{}+
\left(\mu+P_r+2\epsilon-\frac{4}{3}\eta\sigma\right)\frac{\dot B}{B}\right. \nonumber\\
\left.+2\left(\mu+P_{\perp}+\epsilon+\frac{2}{3}\eta\sigma\right)\frac{\dot R}{R}\right] \nonumber\\
-\frac{1}{B}\left[(q+\epsilon)^{\prime}+2(q+\epsilon)\frac{(AR)^{\prime}}{AR}\right]=0, \label{j4}\\
T^{-\alpha\beta}_{;\beta}\chi_{\alpha}=\frac{1}{A}\left[(q+\epsilon)\dot{}
+2(q+\epsilon)\left(\frac{\dot B}{B}+\frac{\dot R}{R}\right)\right] \nonumber\\
+\frac{1}{B}\left[\left(P_r+\epsilon-\frac{4}{3}\eta\sigma\right)^{\prime}
+\left(\mu+P_r+2\epsilon-\frac{4}{3}\eta\sigma\right)\frac{A^{\prime}}{A}\right. \nonumber\\
\left.+2(P_r-P_{\perp}+\epsilon-2\eta\sigma)\frac{R^{\prime}}{R}\right]=0, \label{j5}
\end{eqnarray}
or, by using (\ref{5c}), (\ref{5c1}), (\ref{16}), (\ref{23a}) and (\ref{20x}), they become, respectively,
\begin{eqnarray}
D_T(\mu+\epsilon)+\frac{1}{3}\left(3\mu+P_r+2P_{\perp}+4\epsilon \right)\Theta \nonumber\\
+\frac{2}{3}(P_r-P_{\perp}+\epsilon-2\eta\sigma)\sigma+ED_R(q+\epsilon)
+2(q+\epsilon)\left(a+\frac{E}{R}\right)=0, \label{j6}\\
D_T(q+\epsilon)+\frac{2}{3}(q+\epsilon)(2\Theta+\sigma)
+ED_R\left(P_r+\epsilon-\frac{4}{3}\eta\sigma\right) \nonumber\\
+\left(\mu+P_r+2\epsilon-\frac{4}{3}\eta\sigma\right)a+2(P_r-P_{\perp}+\epsilon-2\eta\sigma)\frac{E}{R}=0.
\label{j7}
\end{eqnarray}
This last equation may be further tranformed as follows: The acceleration $D_TU$ of an infalling particle inside $\Sigma$ can
be obtained by using (\ref{5c}), (\ref{14}), (\ref{17masa})  and (\ref{20x}),
producing
\begin{equation}
D_TU=-\frac{m}{R^2}-4\pi\left(P_r+\epsilon-\frac{4}{3}\eta\sigma\right)R
+Ea, \label{28}
\end{equation}
and then, substituting $a$ from (\ref{28}) into
(\ref{j7}), we obtain
\begin{eqnarray}
\left(\mu+P_r+2\epsilon-\frac{4}{3}\eta\sigma\right)D_TU \nonumber\\
=-\left(\mu+P_r+2\epsilon-\frac{4}{3}\eta\sigma\right)
\left[\frac{m}{R^2}
+4\pi\left(P_r+\epsilon-\frac{4}{3}\eta\sigma\right)R\right] \nonumber\\
-E^2\left[D_R\left(P_r+\epsilon-\frac{4}{3}\eta\sigma\right)
+2(P_r-P_{\perp}+\epsilon-2\eta\sigma)\frac{1}{R}\right] \nonumber\\
-E\left[D_T(q+\epsilon)+2(q+\epsilon)\left(2\frac{U}{R}+\sigma\right)\right].
\label{3m}
\end{eqnarray}
The  physical meaning of different terms in (\ref{3m}) has been discussed in detail in \cite{matter}-\cite{matter2}. Suffice to say in this point that  the first term on the right hand side describes the gravitational force term.

\section{Shearing expansionfree motion}
If the fluid has no expansion, i.e. $\Theta=0$, then from (\ref{5c1}) we have
\begin{equation}
\frac{\dot B}{B}=-2\frac{\dot R}{R}, \label{20}
\end{equation}
or, by integrating
\begin{equation}
B=\frac{g(r)}{R^2}, \label{21}
\end{equation}
where $g(r)$ is an arbitrary function of $r$.

Substituting (\ref{20}) into (\ref{13}) we obtain
\begin{equation}
\frac{{\dot R}^{\prime}}{R}+2\frac{\dot R}{R}\frac{R^{\prime}}{R}-\frac{\dot R}{R}\frac{A^{\prime}}{A}=4\pi(q+\epsilon)AB, \label{22}
\end{equation}
which can be integrated for ${\dot R}\neq 0$ producing
\begin{equation}
A=\frac{R^2{\dot R}}{\tau_1}\exp\left[-4\pi\int(q+\epsilon)AB\frac{R}{\dot R}dr\right], \label{23}
\end{equation}
where $\tau_1(t)$ is an arbitrary function of $t$.
With (\ref{21}) and {\ref{23})
then (\ref{1}) becomes
\begin{eqnarray}
ds^2=-\left\{\frac{R^2 {\dot R}}{\tau_1}\exp\left[
-4\pi\int(q+\epsilon)AB\frac{R}{\dot R}dr\right]\right\}^2dt^2 \nonumber\\
+\left(\frac{g}{R^2}\right)^2dr^2+R^2(d\theta^2+\sin^2\theta d\phi^2),
\label{25II}
\end{eqnarray}
which is the general metric for a shearing expansionfree anistropic dissipative fluid.

In the nondissipative case $q=\epsilon=0$ we can write (\ref{25II}) as
\begin{equation}
ds^2=-\left(\frac{R^2 {\dot R}}{\tau_1}\right)^2dt^2+\frac{1}{R^4}dr^2+R^2(d\theta^2+\sin^2\theta d\phi^2),
\label{25III}
\end{equation}
where  without loss of generality (by reparametrizing $r$) we put $g=1$ (observe that  a unit constant with dimensions $[r^4]$ is assumed to multiply $dr^2$). Then we have that (\ref{25III}) is the general metric for a spherically symmetric
anisotropic perfect fluid undergoing shearing and expansionfree evolution  (observe that it  has the same form as for the isotropic fluid \cite{Mac}).

\section{On the physical meaning of  expansionfree motion}
We shall now analyze under which conditions a time dependent spherically symmetric configuration may evolve without expansion. For doing so we shall try to develop more our understanding of the different speeds that involve the description of expansion as well as shear in the evolution of a selfgravitating fluid. The following  discussion heavily relies on the kinematic quantities characterizing the motion of a medium presented in \cite{dem}, with slight changes in notation.

In Gaussian coordinates, the position of each particle may be given as
\begin{equation}
x^\alpha=x^\alpha(y^a,s),
\label{vel1}
\end{equation}
where $s$ is the proper time along the world line of the particle, and $y^a$ (with $a$ running from 1 to 3) is the position of the particle on a three-dimensional hypersurface (say $\Sigma$).
Then for the unit vector tangent to the world line (the four-velocity) we have
\begin{equation}
V^\alpha=\frac{\partial x^\alpha}{\partial s},
\label{vel2}
\end{equation}
and observe that
\begin{equation}
\frac{\partial }{\partial s}=D_T.
\label{nueva}
\end{equation}
Next, for an infinitesimal variation of the world line we have
\begin{equation}
\delta x^\alpha=\frac{\partial x^\alpha}{\partial y^a} \delta y^a,
\label{vel3}
\end{equation}
from which it follows
\begin{equation}
D_T (\delta x^\alpha)=V^\alpha_{;\beta}\delta x^\beta.
\label{vel4}
\end{equation}
Introducing the projector $h^\alpha_{\beta}$ on $\Sigma$ by
\begin{equation}
h^\alpha_{\beta}=\delta ^\alpha_{\beta}+V^\alpha V_{\beta},
\label{vel5}
\end{equation}
we can define the position vector of the particle $y^a+\delta y^a$ relative to the particle $y^a$ on $\Sigma$, as
\begin{equation}
\delta_{\bot}x^\alpha=h^\alpha_{\beta} \delta x^\beta.
\label{vel6}
\end{equation}
Then the relative velocity between these two particles, is
\begin{equation}
u^\alpha=h^\alpha_{\beta} D_T(\delta_{\bot} x^\beta),
\label{vel7}
\end{equation}
and considering (\ref{vel4}) and  (\ref{vel6}) it follows that
\begin{equation}
u^\alpha=V^\alpha_{;\beta} \delta_{\bot} x^\beta.
\label{vel8}
\end{equation}

Now, the infinitesimal distance between two neighboring points on $\Sigma$ is
\begin{equation}
\delta l^2=g_{\alpha \beta} \delta_{\bot} x^\beta \delta_{\bot} x^\alpha,
\label{vel9}
\end{equation}
then
\begin{equation}
\delta l D_T(\delta l)=g_{\alpha \beta} \delta_{\bot} x^\beta D_T(\delta_{\bot} x^\alpha),
\label{vel10}
\end{equation}
or, by using (\ref{vel4}) and (\ref{vel7}),
\begin{equation}
\delta l D_T(\delta l)=V_{\alpha; \beta} \delta_{\bot} x^\beta \delta_{\bot} x^\alpha.
\label{vel11}
\end{equation}
Then, taking into consideration the expression for the irreducible components of a timelike vector
\begin{equation}
V_{\alpha
;\beta}=\sigma_{\alpha\beta}-a_{\alpha}V_{\beta}+\frac{1}{3}\Theta h_{\alpha\beta},
\label{vel12}
\end{equation}
where we assumed zero rotation, and substituting
into (\ref{vel11})
we obtain
\begin{equation}
\delta l D_T(\delta l)= \delta_{\bot} x^\beta \delta_{\bot} x^\alpha\left(\sigma_{\alpha \beta}+\frac{1}{3}h_{\alpha \beta} \Theta\right),
\label{vel13}
\end{equation}
or, introducing the spacelike unit vector
\begin{equation}
e^\alpha=\frac{\delta_{\bot} x^\alpha}{\delta l}, \label{unit}
\end{equation}
it becomes
\begin{equation}
 \frac{D_T(\delta l)}{\delta l}= e^\alpha e^\beta \sigma_{\alpha \beta}+\frac{\Theta}{3}.
\label{vel14}
\end{equation}

Let us now consider, the spherically  symmetric case, and apply (\ref{vel14}) to two neighbouring points along the radial direction. In this case we have $e^\alpha \equiv \chi^\alpha$,
and using (\ref{5}),  (\ref{5a}) and (\ref{5b}) in (\ref{vel14}) we obtain
\begin{equation}
 \frac{D_T(\delta l)}{\delta l}= \frac{2\sigma}{3}+\frac{\Theta}{3},
\label{vel15}
\end{equation}
or, by using (\ref{5b1}) and (\ref{5c1})
\begin{equation}
 \frac{D_T(\delta l)}{\delta l}= \frac{\dot B}{AB}.
\label{vel16}
\end{equation}
Then with (\ref{19}) and (\ref{vel16}) we can write
\begin{equation}
 \sigma= \frac{D_T(\delta l)}{\delta l}-\frac{D_T R}{R}= \frac{D_T(\delta l)}{\delta l}-\frac{U}{R},
\label{vel17}
\end{equation}
and
\begin{equation}
 \Theta= \frac{D_T(\delta l)}{\delta l}+\frac{2D_T R}{R}=\frac{D_T(\delta l)}{\delta l}+\frac{2U}{R},
\label{vel17bis}
\end{equation}
Thus we see that in general there are two different contributions to the shear (\ref{vel17}) and to the expansion (\ref{vel17bis}). One is due to the ``circumferential'' velocity $U$ \cite{ray}, which is  related to the change  of areal radius $R$  of a layer of matter, whereas the other is related to
$D_T(\delta l)$, which has also the meaning of ``velocity'', being the relative velocity between neighboring layers of matter, and can be in general different from $U$.

From (\ref{vel17}) we see that, if the spherical distribution of matter is collapsing, $U<0$, the shear can vanish only if the relative distance between different layers of matter diminishes, $D_T(\delta l)<0$, and cancels the circumferential velocity.

From (\ref{vel17bis}) we see that  the evolution of the fluid will be expansionfree, whenever the ``circumferential''  term cancels the term related to the variation of distance of neighboring particles. Thus the collapse will proceed expansionfree, if  the decrease of the perimeter of a comoving sphere ($U<0$) is compensated by an increase in the distance of neighbouring particles (along the radial direction)  according to (\ref{vel17bis}). Alternatively,  if the fluid is moving outward ($U>0$) neghbouring particles  will get closer ($D_T(\delta l)<0$). These observations clarify further the origin of the cavity in  expansionfree models. Indeed, consider two concentric fluid shells in the neighborhood of the centre. As it follows from (\ref{26}), close to the centre we have $U\sim R$. Now, in the process of expansion (increasing of $R$), the $\Theta=0$ condition implies as mentioned before that 
$D_T(\delta l)<0$, i.e. both shells become closer, however this would not be so as long as $U\sim R$, implying thereby that the $\Theta=0$ condition requires that the innermost  shell of fluid should be away from the centre,  initiating therefrom  the formation of the cavity.

Let us see this from another perspective. Consider the infinitesimal volume of the shell  between two concentric spheres of radii $r$ and $r+\delta r$,
\begin{equation}
\delta V=4\pi BR^2\delta r,
\label{vel18}
\end{equation}
then it follows
\begin{equation}
D_T(\delta V)=4\pi (D_TB)R^2\delta r+8\pi B R(D_TR)\delta r,
\label{vel19}
\end{equation}
or, dividing (\ref{vel19}) by $\delta V$ and using (\ref{vel16}),
\begin{equation}
\frac{D_T(\delta V)}{\delta V}=\frac{D_T(\delta l)}{\delta l}+2 \frac{U}{R}.
\label{vel20}
\end{equation}
Which of course coincide with (\ref{vel17bis}) since we know that $D_T(\delta V)/\delta V$ is the definition of the expansion. \

Thus in the process of contraction
(expansion), the elementary volume $\delta V$ decreases (increases) by two factors. On the one hand by the decreasing (increasing) of the areal radius $R$ and on the other hand by
the decreasing (increasing) of the proper radial distance between the two concentric surfaces.  Again, we have an expansionfree evolution (the elementary volume $\delta V$ remains constant) whenever the two contributions on the right hand of  (\ref{vel20}) cancel each other, in spite of the fact that neither of them vanishes.

We shall now see how these two different definitions of radial velocity considered above are related.

Let us first  assume that $U=0$. Then from (\ref{vel17}) and (\ref{vel17bis}) it follows $\Theta=\sigma$, feeding this back into
(\ref{21a}) we get at once
\begin{equation}
D_T(\delta l)=-\frac{4\pi R(q+\epsilon)}{E}\delta l,
\label{relv1}
\end{equation}
thus, $U=0$ implies $D_T( \delta l) =0$ only in the dissipationless case, $q=\epsilon=0$.

Next, let us assume $D_T(\delta l)=0$. Then it follows from (\ref{vel17}) and (\ref{vel17bis}) that $\Theta=-2\sigma$, feeding this  back into
(\ref{21a}) we get
\begin{equation}
D_R\sigma+\frac{\sigma}{R}=-\frac{4\pi (q+\epsilon)}{E}
\label{27}
\end{equation}
whose integration with respect to $R$ yields
\begin{equation}
\sigma=\frac{\zeta}{R}-\frac{4\pi}{R}\int(q+\epsilon)\frac{R}{E}dR,
\label{28relvel}
\end{equation}
where $\zeta$ is independent of $R$ (for any layer of fluid, characterized by $r=constant$,  $\zeta$ may depend on $t$. In general however it may depend on $t$ and $r$). On the other hand, (\ref{28relvel}) with (\ref{vel17}) implies
\begin{equation}
U=-\zeta+4\pi\int(q+\epsilon)\frac{R}{E}dR.
\label{29relvel}
\end{equation}
Since $U\rightarrow 0$ as $R \rightarrow 0$, we must put $\zeta=0$ (if only the centre of the fluid distribution is covered by  the coordinate system). Thus, in the nondissipative case $U=0$.

Therefore only in the nondissipative case $U=0 \leftrightarrow D_T(\delta l)=0$ (with the condition mentioned above).

Finally, we can write (\ref{21a}) as
\begin{equation}
D_R\left(\frac{U}{R}\right)=\frac{4 \pi}{E}(q+\epsilon)+\frac{\sigma}{R}
\label{vel24}
\end{equation}
which after integration with respect to $R$ becomes
\begin{equation}
U=\xi R+R\int^R_0\left[\frac{4\pi}{E}(q+\epsilon)+\frac{\sigma}{R}\right]dR,
\label{25}
\end{equation}
or,
\begin{equation}
U=\frac{U_{\Sigma^{(e)}}}{R_{\Sigma^{(e)}}}R-R\int^{R_{\Sigma^{(e)}}}_R\left[\frac{4\pi}{E}(q+\epsilon)+\frac{\sigma}{\tilde R}\right]d\tilde R.
\label{26}
\end{equation}
In the shearfree nondissipative case we have from (\ref{26}) that $U\sim R$, which is characteristic of the homologous evolution \cite{7'}.
This implies that for two concentric shells of areal radii $R_1$ and $R_2$, we have in this case
\begin{equation}
\frac{R_1}{R_2}=\mbox{constant}.
\label{vel22}
\end{equation}
The second term on the right of (\ref{26}) describes how the shear and dissipation deviate the evolution from the homologous regime.

It is worth noticing that in the shearfree nondissipative case the sign of $U$ for any fluid element is the same as that of  $U_{\Sigma^{(e)}}$. However if  we relax any of those conditions, that might not   be true. Thus it would be possible, for example, to have an expandig outer shell  with an  imploding inner core. Such  possibility was brought out before, but restricted  to  the quasistatic regime \cite{two}. Here we see that  such an scenario is also possible in the general dynamic regime.

\section{Shearing expansionfree perfect fluid}
We shall now  restrict our study to a shearing expansionfree fluid without dissipation, $q=\epsilon=0$ and $\eta=0$, then the metric reduces to (\ref{25III}) and the field equations (\ref{12}-\ref{15}), using (\ref{21a}), become
\begin{eqnarray}
8\pi\mu=-2R^3R^{\prime\prime}-5R^2R^{\prime 2}+\frac{1}{R^2}-3\frac{\tau^2_1}{R^6}, \label{40}\\
\frac{1}{3}D_R \sigma+\frac{\sigma}{R}=0,\label{21aexf}\\
8\pi P_r=\frac{\tau^2_1}{R^5{\dot R}}\left(3\frac{\dot R}{R}-2\frac{{\dot \tau}_1}{\tau_1}\right)
+R^3R^{\prime}\left(2\frac{{\dot R}^{\prime}}{\dot R}+5\frac{R^{\prime}}{R}\right)-\frac{1}{R^2}, \label{41}\\
8\pi P_{\perp}=-\frac{\tau^2_1}{R^5{\dot R}}\left(6\frac{\dot R}{R}-\frac{{\dot \tau}_1}{\tau_1}\right)
+R^4\left[\frac{{\dot R}^{\prime\prime}}{\dot R}
+7\frac{{\dot R}^{\prime}}{\dot R}\frac{R^{\prime}}{R}+3\frac{R^{\prime\prime}}{R}
+10\left(\frac{R^{\prime}}{R}\right)^2\right]; \label{42}
\end{eqnarray}
while the Bianchi identities (\ref{j4}) and ({\ref{j5}) read,
\begin{eqnarray}
{\dot\mu}+2(P_{\perp}-P_r)\frac{\dot R}{R}=0, \label{43}\\
P_r^{\prime}+(\mu+P_r)\frac{{\dot R}^{\prime}}{\dot R}+2(\mu+2P_r-P_{\perp})\frac{R^{\prime}}{R}=0. \label{44}
\end{eqnarray}
From (\ref{43}) we have that if the fluid is isotropic, $P_r=P_{\perp}$, then the energy density $\mu$ is only $r$ dependent.

We can now integrate (\ref{40}) under the assumption $\mu=\mu(r)$, to obtain
\begin{equation}
R^{\prime 2}=\frac{1}{R^4}+\frac{\tau_2-2m}{R^5}+\frac{\tau^2_1}{R^8}, \label{45I}
\end{equation}
where $\tau_2(t)$ is an arbitrary function of $t$ and (\ref{27intcopy}) has been used.

We shall now specialize further our model to the case of constant energy density \cite{Skripkin}

\subsection{The Skripkin model}
In \cite{Skripkin} it is not explicitly  assumed that $\Theta=0$, instead it is assumed   that the fluid is nondissipative, has its energy density $\mu=\mu_0=$ constant  and the pressure isotropic. Of course, these conditions imply, because of (\ref{j6}), that $\Theta=0$. Thus we have only  one physical variable ($P_r$) and the system of field equations is closed and can be integrated.

From the condition $\mu=\mu_0=$ constant, (\ref{45I}) becomes
\begin{equation}
R^{\prime 2}=-\frac{k}{R^2}+\frac{1}{R^4}+\frac{\tau_2}{R^5}+\frac{\tau^2_1}{R^8}, \label{45}
\end{equation}
with
\begin{equation}
k=\frac{8\pi\mu_0}{3}. \label{46}
\end{equation}

It should be observed that (\ref{45}) imposes a maximum to the value of $R$ ($R_{max}$),  for which $R^{\prime}=0$. The physical origin of this maximum for the areal radius may be explained as follows:

In the Skripkin picture the fluid is initially at rest, then there is a   sudden explosion at  the centre producing the outward ejection of the fluid, always  keeping the conditions of nondissipation, $\mu=\mu_0=$ constant  and isotropic  pressure (i.e. $\Theta=0$). Under these  conditions, (\ref{3m}) becomes
\begin{equation}
\left(\mu+P\right) D_TU =-\left(\mu+P\right)\left(\frac{m}{R^2}
+4\pi P R\right) -E^2D_R P,
\label{3mbis}
\end{equation}
with $P_r=P_{\perp}=P$. Now, as the fluid moves outward and $R$  approaches $R_{max}$, $R^{\prime}$ tends to zero which implies, because of (\ref{20x}), that $E$ approaches zero. Furthermore, this implies that the gravitational (negative) term in (\ref{3mbis}) will prevail leading to a negative $D_TU$, producing a reversal of the motion at (or before) $R_{max}$.

As mentioned above, Skripkin assumes the pressure to be isotropic.  However if we assume  the evolution to be expansionfree (and $\mu=\mu_0=$ constant) then the isotropy of pressure follows  from (\ref{43}), and using (\ref{45}) in (\ref{41}) or
(\ref{42}) yields
\begin{equation}
8\pi P=\frac{{\dot \tau}_2}{R^2{\dot R}}-3k, \label{47}
\end{equation}
which, of course, satisfies (\ref{44}). From the matching condition (\ref{j3}) we have for (\ref{47})
\begin{equation}
8\pi P\stackrel{\Sigma^{(e)}}{=}\frac{{\dot\tau}_2}{R^2{\dot R}}-3k=0. \label{48}
\end{equation}
which gives
\begin{equation}
\tau_2=kR^3_{\Sigma^{(e)}}+c_1, \label{49}
\end{equation}
where $c_1$ is an arbitrary constant.

The mass function (\ref{17masa}) with (\ref{25III}) and (\ref{45}) becomes
\begin{equation}
m=\frac{1}{2}(kR^3-\tau_2)=\frac{k}{2}(R^3-R^3_{\Sigma^{(e)}})-\frac{c_1}{2}, \label{50}
\end{equation}
where we used (\ref{49}).
Measuring $m$ on $\Sigma^{(e)}$ we obtain the total mass  of the configuration $M$
\begin{equation}
m\stackrel{\Sigma^{(e)}}{=}M=-\frac{c_1}{2}. \label{51}
\end{equation}
Thus
\begin{equation}
\tau_2=kR^3_{\Sigma^{(e)}}-2M.\label{52}
\end{equation}
and
\begin{equation}
m=\frac{k}{2}(R^3-R^3_{\Sigma^{(e)}})+M. \label{53}
\end{equation}

As mentioned before, it should be clear from physical considerations that the assumption of  vanishing expansion  (with the constant energy density  condition)  in the evolution of the fluid distribution, implies the formation of a vacuum cavity  within the sphere.

Applying matching conditions (\ref{junction1i}) and (\ref{j3i}) on the boundary surface $\Sigma^{(i)}$, delimiting the cavity, we obtain
\begin{equation}
M=\frac{k}{2}(R^3_{\Sigma^{(e)}}-R^3_{\Sigma^{(i)}}), \label{55}
\end{equation}
and using (\ref{55}) in (\ref{52})
\begin{equation}
\tau_2=kR^3_{\Sigma^{(i)}}.\label{52bisI}
\end{equation}

Due to the constancy of $M$ we obtain from  (\ref{55})
\begin{equation}
\dot R_{\Sigma^{(e)}}=\left(\frac{R_{\Sigma^{(i)}}}{R_{\Sigma^{(e)}}}\right)^2\dot R_{\Sigma^{(i)}}, \label{56}
\end{equation}
and from   (\ref{25III}) and (\ref{56})
\begin{equation}
A_{\Sigma^{(e)}}=A_{\Sigma^{(i)}},
\label{ab}
\end{equation}
producing,  because of (\ref{56}),
\begin{equation}
U_{\Sigma^{(e)}}=\left(\frac{R_{\Sigma^{(i)}}}{R_{\Sigma^{(e)}}}\right)^2U_{\Sigma^{(i)}},
\label{Ub}
\end{equation}
which implies, as expected, that the inner boundary surface $\Sigma^{(i)}$ moves faster  than the outer  boundary surface $\Sigma^{(e)}$. This result can also be deduced from the very definition of $U$.

Indeed, using (\ref{25III}) in (\ref{19}) we obtain
\begin{equation}
U=\frac{\tau_1}{R^2},
\label{uexpf}
\end{equation}
which evaluated on $\Sigma^{(i)}$ and $\Sigma^{(e)}$ produces
\begin{equation}
\tau_1=U_{\Sigma^{(e)}}R_{\Sigma^{(e)}}^2=U_{\Sigma^{(i)}}R_{\Sigma^{(i)}}^2,
\label{otrou}
\end{equation}
implying (\ref{Ub}).

Observe that from (\ref{vel17}), (\ref{vel17bis}) and (\ref{uexpf}) it follows
\begin{equation}
\sigma=-\frac{3\tau_1}{R^3},
\label{sigmaexf}
\end {equation}
which is the solution of (\ref{21aexf}).

It should be observed that since the pressure vanishes on $\Sigma^{(i)}$ and $\Sigma^{(e)}$, it should have a maximum somewhere between the two, i.e the pressure gradient must vanish  on some spherical surface  ($S$) within the fluid. If we denote the areal radius of such surface  by $R=R_S$, then it follows from (\ref{44})
\begin{equation}
(R^2R^{\prime})^{\dot{}}\stackrel{S}{=}0 \label{44new}
\end{equation}
and after integration
\begin{equation}
R^{\prime}\stackrel{S}{=}\frac{c_2}{R^2}
\label{55new}
\end{equation}
where $c_2$ is a constant. Substituting (\ref{55new}) into (\ref{45}) we have
\begin{equation}
kR^6+(c_2^2-1)R^4-\tau_2R^3-\tau_1^2\stackrel{S}{=}0. \label{66new}
\end{equation}
Thus the surface $R_S$, which is the root of (\ref{66new}), divides the fluid into two regions. The inner one, with a positive pressure gradient, and the outer one with a negative pressure gradient.

We are  now able to prescribe the strategy to determine the Skripkin models. First of all let us recall that  without loss of generality  Skripkin chooses $\tau_1=R_{\Sigma^{(e)}}  \dot R_{\Sigma^{(e)}}^2$ and, consequently,  $\tau_1=R_{\Sigma^{(i)}}  \dot R_{\Sigma^{(i)}}^2$ and $A_{\Sigma^{(e)}}=A_{\Sigma^{(i)}}=1$.

Then, the integration of  (\ref{45}), which can only be expressed in terms of elliptic functions, produces
\begin{equation}
R=R(r,R_{\Sigma^{(i)}},\dot R_{\Sigma^{(i)}}).
\label{st1}
\end{equation}
Evaluating (\ref{3mbis}) on $\Sigma^{(i)}$ we obtain a differential equation for $R_{\Sigma^{(i)}}$ whose integration  provides its  time dependence, and therefore of $R(r,t)$.

If we deviate from Skripkin model and relax the condition $\mu=$ constant, allowing for $r$ dependence of $\mu$, then we need to integrate (\ref{45I}) instead of (\ref{45}), which of course requires the specific $r$ dependence of $\mu$ or $m$.

Also, we could consider the anisotropic case, which allows for a $t$ dependence of $\mu$, in this case, of course, a specific equation of state for the anisotropic pressures is required  (or an equivalent ansatz).

Finally the integration in the general dissipative case would require a thorough knowledge of the energy production within the fluid (or a set of equivalent ans\"atze).

\section{Conclusions}
We have seen so far that  expansionfree condition allows for the obtention  of a wide range of models for  the evolution of spherically symmetric  selfgravitating systems. Ranging from  nondissipative spheres, with constant energy density and isotropic pressure (Skripkin model), to  general dissipative anisotropic models.

Observe that even if the Skripkin model is the simplest, from the physical point of view, it might not be so from the mathematical point of view.  Indeed, we could in principle choose a mass function, such  that  (\ref{45I}) could be integrated in terms of elementary functions, obviously it remains to be seen if such  models  are  endowed with any physical interest.

One of the most interesting features of the models, is the appearance of a vacuum cavity  within the fluid distribution. It is not clear at this point if such models  might be used to describe the formation of voids observed at cosmological scales (see \cite{voids} and references therein).

The two  concepts of radial velocity discussed in section 5 allows to understand the meaning of the expansionfree evolution. As a by-product of such discussion, the  shearfree flow (in the nondissipative case)  appears to be equivalent to the well known homologous evolution. Particularly remarkable is the fact  that the expansionfree ejection (collapse) implying an increase (decrease) in the areal radius of a layer of matter, proceeds with a decrease (increase) in the distance of neighbouring particles along the radial direction. Also, the possibility  of  a ``splitting'' of the fluid distribution (change of sign in $U$) due to dissipation  and/or shear, as indicated by (\ref{26}), deserves to be explored further.

Finally, it is worth noticing that in the locally anisotropic case, the expansionfree evolution, due to the second term on the left of equation (\ref{43}), does not imply  that energy density remains time independent. This situation becomes intelligible when  it is remembered, that in the limit of hydrostatic
equilibrium, when $U=q=\epsilon=0$, we obtain from (\ref{j7}),
\begin{equation}
D_RP_r+\frac{2(P_r-P_{\perp})}{R}=-\frac{\mu+P_r}{R(R-2m)}
\left(m+4\pi P_rR^3\right),
\label{30int}
\end{equation}
which is just the generalization of the Tolman-Oppenheimer-Volkoff equation for anisotropic fluids, obtained in comoving coordinates  \cite{Chan}. Thus the term $2(P_r-P_{\perp})/R$ represents a force associated to the local anisotropy of pressure, and therefore the second term on the left  of equation (\ref{43}), is the rate of work done by that force, resulting in a  change of $\mu$.

\section*{Acknowledgments.}
LH  acknowledges financial support from the CDCH at Universidad Central
de Venezuela under grants PG 03-00-6497-2007 and PI 03-00-7096-2008.

\end{document}